\documentclass[sigconf]{acmart}
\AtBeginDocument{%
  \providecommand\BibTeX{{%
    \normalfont B\kern-0.5em{\scshape i\kern-0.25em b}\kern-0.8em\TeX}}}

\copyrightyear{2024} 
\acmYear{2024} 
\setcopyright{rightsretained} 
\acmConference[WWW '24 Companion]{Companion Proceedings of the ACM Web Conference 2024}{May 13--17, 2024}{Singapore, Singapore}
\acmBooktitle{Companion Proceedings of the ACM Web Conference 2024 (WWW '24 Companion), May 13--17, 2024, Singapore, Singapore}\acmDOI{10.1145/3589335.3651556}
\acmISBN{979-8-4007-0172-6/24/05}

\usepackage{multirow}
\usepackage[normalem]{ulem}
\useunder{\uline}{\ul}{}
\usepackage{bbm}
\usepackage{enumitem}
\usepackage{tablefootnote}
\usepackage{makecell}
\usepackage[ruled,linesnumbered,longend]{algorithm2e}
\usepackage{algpseudocode}  
\usepackage{amsmath}

\begin{document}

\title[Rumor Mitigation in Social Media Platforms with Deep Reinforcement Learning]{Rumor Mitigation in Social Media Platforms with \\Deep Reinforcement Learning}

\author{Hongyuan Su}
\affiliation{%
  \institution{Department of Electronic
Engineering, BNRist, 
Tsinghua University}
  \city{Beijing}
  \country{China}}

\author{Yu Zheng}
\affiliation{%
  \institution{Department of Electronic
Engineering, BNRist, 
Tsinghua University}
  \city{Beijing}
  \country{China}}

\author{Jingtao Ding}
\affiliation{%
  \institution{Department of Electronic
Engineering, BNRist, 
Tsinghua University}
  \city{Beijing}
  \country{China}}

\author{Depeng Jin}
\affiliation{%
  \institution{Department of Electronic
Engineering, BNRist, 
Tsinghua University}
  \city{Beijing}
  \country{China}}

\author{Yong Li}
\authornote{Corresponding author (liyong07@tsinghua.edu.cn).}
\affiliation{%
  \institution{Department of Electronic
Engineering, BNRist, 
Tsinghua University}
  \city{Beijing}
  \country{China}}

\renewcommand{\shortauthors}{Hongyuan Su, Yu Zheng, Jingtao Ding, Depeng Jin, \&Yong Li}

\begin{abstract}

Social media platforms have become one of the main channels where people disseminate and acquire information, of which the reliability is severely threatened by rumors widespread in the network.
Existing approaches such as suspending users or broadcasting real information to combat rumors are either with high cost or disturbing users.
In this paper, we introduce a novel rumor mitigation paradigm, where only a minimal set of links in the social network are intervened to decelerate the propagation of rumors, countering misinformation with low business cost and user awareness.
A knowledge-informed agent embodying rumor propagation mechanisms is developed, which intervenes the social network with a graph neural network for capturing information flow in the social media platforms and a policy network for selecting links.
Experiments on real social media platforms demonstrate that the proposed approach can effectively alleviate the influence of rumors, substantially reducing the affected populations by over 25\%.
Codes for this paper are released at \textcolor{blue}{\url{https://github.com/tsinghua-fib-lab/DRL-Rumor-Mitigation}}.

\end{abstract}



\maketitle

\section{Introduction}

The rapid spread of rumors emerges as an unprecedented challenge, causing detrimental impact to public safety, health, and democracy.
Central to the rapid dissemination of misinformation is the phenomenon of information cascades, where social network users influenced by the perceived information, adopt or share that information, leading to a cascading effect triggered and manipulated by strategically placed misinformation~\cite{ding2024artificial,ding2019reinforced}.
Therefore, it is critical to develop an intelligent algorithm that can automatically synthesize rumor mitigation actions to counter the large number of rumors emerging everyday across the social network.

Current approaches for mitigating rumors follow two major paradigms.
First, a subset of users are selected and suspended, \textit{i.e.} removing nodes from the network to block the propagation of rumors~\cite{meirom2021controlling}.
Second, real information is introduced to the network, where several seed nodes are chosen for broadcasting real information to maximize its influence over the network~\cite{budak2011limiting}.
However, these methods tend to be costly and cause significant annoyance to users.
Therefore, in this paper, we focus on a new paradigm where rumors are mitigated with low business cost, and users are almost unaware of the mitigation actions~\cite{yan2019rumor}.
We reserve all nodes and only block a small fraction of \textit{links} in the social network to decelerate the information cascade of rumors, which induces minimal user disturbance and keeps information integrity.

Rumor mitigation via link deletion faces three main challenges.
\textit{First}, the candidate intervention options are extremely large as there are much more links than nodes in a social network, resulting in an enormous solution space.
\textit{Second}, it is difficult to capture the underlying mechanisms that drive the propagation of rumors due to its complexity and randomness.
Thus, how to effectively leverage the domain knowledge of social networks remains largely unexplored by existing literature.
\textit{Third}, real-world social networks tend to exhibit diverse topologies and dynamics~\cite{quan2023robust}.
Consequently, it is necessary to develop an algorithm with sufficient generalization ability, to make it applicable in real rumor mitigation tasks.

To address these challenges, we develop a deep reinforcement learning (DRL) framework for rumor mitigation.
To utilize the rich domain knowledge, we first extract critical features according to social network theories that capture the contribution of each node and edge in rumor propagation.
We then design a graph neural network (GNN) with link-route-aware message passing to capture local and global roles of links, respectively.
A rumor-oriented policy network is further developed to select links for deletion, taking both rumor source and community structures into consideration.
Our knowledge-informed agent embodies rumor propagation mechanisms, effectively promoting the accuracy and explainability of the generated rumor mitigation solutions.
Finally, we design a randomized training environment that simulates social networks with diverse topologies and rumor propagation patterns, which guarantees the generalization ability of our approach.

To summarize, the contributions of this paper are as follows,

\begin{itemize}[leftmargin=*]
    \item We tackle the problem of rumor mitigation in a new paradigm with low cost and user disturbance, which only intervenes a minimal fraction of links of the social network.
    \item We develop a knowledge-informed agent and a randomized training algorithm to achieve generalizable rumor mitigation.
    \item Experiments on real-world social networks verify that our method reduces the influence of rumors substantially by over 25\%.
\end{itemize}

\section{Problem Statement}\label{sec::prob}

Consider a directed social network $\mathcal{G} = (\mathcal{N}, \mathcal{E})$, where $\mathcal{N}$ and $\mathcal{E}$ represent users and influence relationships (i.e. followers and followees) between them.
Assume a rumor initiates from a source $s \in \mathcal{N}$, spreading through a diffusion model $\mu$ that leads to $k$ nodes in the social network receiving the rumor.
The impact of rumor can be defined as the percentage of the population affected,
\begin{equation}
    \eta^{\mu}(\mathcal{G},s) = \dfrac{\mathbb{E}_k[k\,|\,\mathcal{G},s,\mu]}{|\mathcal{N}|},
\end{equation}
where $\mathbb{E}[x]$ denotes the expectation of the random variable $x$, and $|\mathcal{N}|$ represents the number of elements in the set $\mathcal{N}$.
Rumor mitigation can be defined as selecting a subset of edges in the graph for removal to minimize the spread of rumors.
The optimal edge set for deletion can be formulated as: 
\begin{equation}
    \mathcal{E}_D^{*} = \mathop{\arg\min}\limits_{\mathcal{E}_D \subseteq \mathcal{E}, |\mathcal{E}_D|=d} \eta^{\mu}(\mathcal{G}\, \backslash \, \mathcal{E}_D, s),
\end{equation}
where $\mathcal{E}_D$ represents the deleted edge set.

\section{Method}

Figure \ref{fig::pipeline} illustrates the overview of the proposed framework where an agent sequentially removes edges on social networks to prevent rumors from spreading.
A knowledge-informed model is designed to achieve decent edge selection, which encodes rich propagation-related features with a link-route-aware GNN model taking both rumor source and propagation dynamics into consideration.
A rumor-oriented policy network is designed for edge-ranking, perceiving the hierarchical structure and the association with rumor source, enabling high-quality edge blocking for rumor mitigation.
Moreover, we develop a randomized RL training approach to enhance the generalization ability of our model, addressing the challenge of diverse topological structure and uncertain rumor sources.

\subsection{Knowledge-informed Edge Selection }

\noindent{\textbf{Features for Topology and Propagation.}}
Rumor spreading involves multiple aspects, such as rumor source, network structure, and the propagation mechanism.
We design rich topological and rumor spreading related features for nodes (FN) and edges (FE) in social networks, which contain the prior knowledge of rumor spreading and can be categorized into two types.
\begin{itemize}[leftmargin=*]
    \item \textit{Topological features} include degree (FN1, FE6), centrality (FN2), and diffusion importance (FE7)~\cite{liu2015improving}, describing the topological structure of the social network and providing valuable insights into analyzing the spread of rumors.
    For example, diffusion importance indicates the efficiency of information propagation via edges, thus blocking edges with higher diffusion importance can efficiently prevent rumors from reaching more users.
    \item \textit{Rumor propagation features} include shortest path length to rumor source (FN4), rumor source or not (FN5), and betweenness centrality (FN3, FE8) which are closely related to the characteristics of rumor propagation.
    It is evident that users close to the source tend to be affected by the rumor earlier.
    Therefore, edges closer to the source are expected to have a higher removal priority for rumor mitigation.
\end{itemize}
Notably, edge removal alters the topological structure of social network, thus these features are dynamically changing at each step.

\begin{figure}[t]
    \centering
    \includegraphics[width=0.9\linewidth]{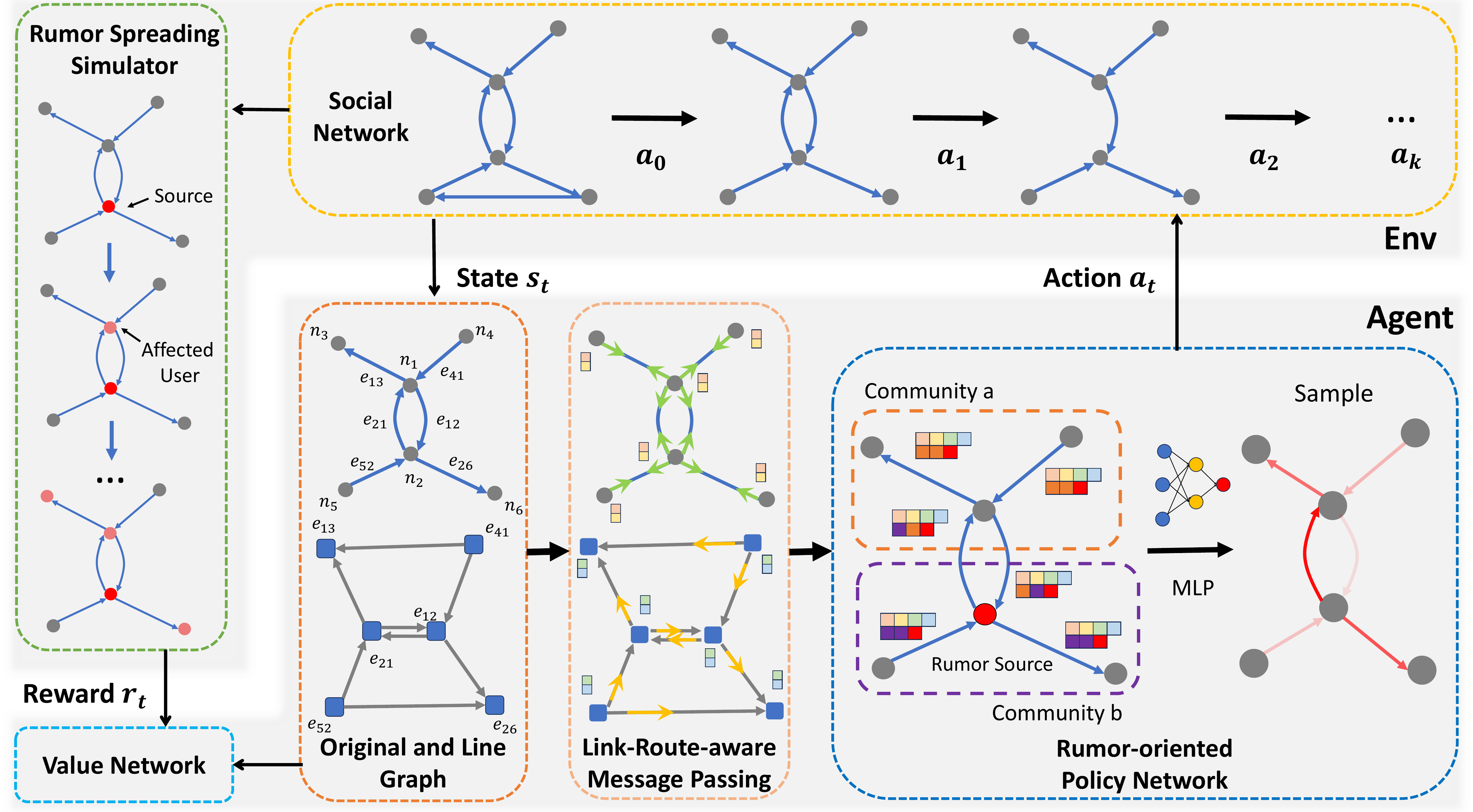}
    \vspace{-5px}
    \caption{
    (Top) The environment provides observations of social network, controls the network based on the actions chosen by the agent, and returns the rewards through the rumor propagation simulations.
    (Bottom) Knowledge-driven edge selection which aggregates neighbors' information through link-route-aware massage passing on both the original and the line graph using a GNN, and finally selects edges with the rumor-oriented policy network.
    }
    \vspace{-15px}
    \label{fig::pipeline}
\end{figure}

\noindent{\textbf{Link-Route-aware Message Passing.}}
Since edges represent influence relationships and serve as key components in rumor propagation, it is necessary to consider their multiple roles as both local user links and global propagation routes.
Therefore, we propose a link-route-aware GNN to learn comprehensive edge embeddings, taking both aspects into consideration, as shown in Figure \ref{fig::pipeline}.
We compute representations for nodes and edges within both the original and the dual line graph as follows,
\begin{align}
    &\mathbf{n}_i^{0} = \mathbf{W}_n^{0}\mathbf{A}_{{n}_i},\quad
    \mathbf{e}_{ij}^{0} = \mathbf{W}_e^{0}\mathbf{A}_{{e}_{ij}},\\
    &\mathbf{n}_{i}^{l+1} = \mathbf{n}_{i}^{l} + \tanh{(\sum_{\mathbf{e}_{ij}\in \mathcal{E}, \mathbf{e}_{ji}\in \mathcal{E}} \mathbf{W}_{n}^{l+1}\mathbf{n}_j^{l})},\\
    &\mathbf{e}_{ij}^{l+1} = \mathbf{n}_q^{*l+1} =  \mathbf{n}_q^{*l} + \tanh{(\sum_{{e}_{pq}^*\in \mathcal{E}^*} \mathbf{W}_{e}^{l+1}\mathbf{n}_p^{*l})},\\
    &\mathbf{e}_{ij}^{f} = [ \mathbf{n}_i^{L} \,\|\, \mathbf{n}_j^{L} \,\|\, \mathbf{e}_{ij}^{L} ],
\end{align}
where $\mathbf{A}_{{n}_i}$ and $\mathbf{A}_{{e}_{ij}}$ are input attributes for nodes and edges, $\mathbf{W}_n^k$ and $\mathbf{W}_e^k$ are linear layers.
Node and edge embeddings are obtained from primal and induced dual line graph, respectively, where edges on primal graph correspond to nodes on line graph, and the asterisk indicates elements in line graph.
The output $\mathbf{n}_i^{L}$ and $\mathbf{e}_{ij}^{L}$ of the last propagation layer are the embeddings of nodes and edges, where $L$ is a hyperparameter and $\|$ means concatenation.

By combining associated two node embeddings of each edge,  $\mathbf{n}_i^{L}$ and  $\mathbf{n}_j^{L}$, we can capture the influence relationships of local user links.
Meanwhile, the edge embedding on the line graph, $\mathbf{e}_{ij}^{L}$, captures the global propagation route role of each edge.
Thus the link-route-aware message passing effectively encode the roles of edges in connecting users and spreading rumors, providing valuable insights for the following edge selection.

\noindent{\textbf{Rumor-oriented Policy Network.}}
We develop a rumor-oriented policy network that takes into account the edge embeddings from GNN, the rumor source, and the communities of social networks.
As illustrated in Figure \ref{fig::pipeline}, the policy network calculates the scores of edges through a multi-layer perceptron (MLP) as follows,
\begin{equation}
    \mathbf{C}_i =  \dfrac{1}{|\mathcal{C}_i|}\sum_{{n}_k\in \mathcal{C}_i} \mathbf{n}_{k}^{L},\quad
    s_{ij} = \verb|MLP|_{p}([\mathbf{e}_{ij}^{f}\,\|\, \mathbf{C}_i \,\|\, \mathbf{C}_j \,\|\, \mathbf{n}_s^{L} ]),\label{eq::logit}
\end{equation}
where $\mathbf{C}_i$ donates the community that node $i$ belongs to, and $\mathbf{n}_s$ donates the embedding of rumor source calculated by GNN.
The agent then samples edges based on probabilities obtained by normalizing the edge scores from the policy network.

\subsection{Generalized Model Training}

As real-world social media platforms exhibit diverse characteristics in network topology and propagation dynamics, it is expected that a rumor mitigation algorithm to be generalizable across different scenarios.
To achieve this, we introduce a randomized RL framework that allows the agent to learn effective rumor mitigation strategies from a large and diverse set of rumor propagation tasks.
In specific, we train the agent with random network topology, where in each episode the environment generates a different social network.
Meanwhile, for the same social network, we simulate the spread of rumors across various topological structures by randomly selecting different nodes as the rumor source at each training episode, which massively increase the number of training social networks.

\section{Experiments}

\subsection{Experiment Settings}

\begin{table}[t]
\small
\caption{
Basic properties of experimented online social networks.
The Nodes and Edges columns indicate the maximum (average) number of corresponding elements in graph.}
\vspace{-10px}
\label{tab::data}
\begin{tabular}{cccccc}
\toprule
\textbf{Network} & \textbf{Type} & \textbf{Nodes} & \textbf{Edges} & \textbf{Diameter} \\
\midrule
Twitter S & Directed & 222\,(78) &  996\,(528) & 10 \\
Twitter M & Directed & 250\,(171) & 7990\,(3086) & 9 \\
Twitter L & Directed & 248\,(215) & 17930\,(9847) & 7 \\
Facebook L & Undirected & 1045\,(417) & 60050\,(17018) & 11 \\
\bottomrule
\end{tabular}
\vspace{-12px}
\end{table}

\noindent{\textbf{Social networks Data.}}
We conduct experiments based on the real-world online social networks including Twitter and Facebook from \href{http://snap.stanford.edu/data/#socnets}{\textit{Stanford Large Network Dataset Collection}}~\cite{leskovec2012learning}.
Table \ref{tab::data} summarizes the basic properties of the adopted datasets.
For each network, we apply the susceptible-infectious-recovered (SIR) model~\cite{hethcote2000mathematics} to simulate rumor propagation with $\gamma=0.20$ and $\beta=0.08$\cite{yu2021modeling}.

\noindent{\textbf{Baselines and Evaluation.}}
We include \textit{traditional approaches} such as Pagerank (PR)~\cite{page1998pagerank}, K-EDGEDELETION (KED)~\cite{tong2012gelling}, and Greedy with Bond Percolation (GBP)~\cite{kimura2008minimizing}.
We also compare our model with \textit{heuristic search} approaches based on degree (HSD) and betweenness (HSB), and \textit{evolutionary algorithms} including genetic algorithm (GA)~\cite{parimi2021genetic} and simulated annealing (SA)~\cite{li2017positive}.
Moreover, we include a recent DRL baseline~\cite{meirom2021controlling} which selects nodes (DRLN) based on embeddings calculated from GNN.

\subsection{Overall Performance}\label{sec::perf_comp}

We remove 10\% of the influencing relationships in the social network, while retaining at least 60\% of the influencing relationships for each user.
Table \ref{tab::overall} illustrates the results of our model and baselines, and we have the following observations,
\begin{itemize}[leftmargin=*]
    \item \textbf{DRL-based methods have significant advantages over other baselines.}
    The DRL-based methods outperform HSC, KED and greedy method on diverse social networks.
    Specifically, they exceed Greedy method with an average improvement of 10.9\%, 9.5\%, and 19.5\% in small, medium and large-sized social networks, respectively.
    Moreover, the DRL-based methods can be easily applied to complex social networks with stable outcome and reasonable computational cost.

    \item \textbf{Our proposed model achieves the best performance.}
    Our approach maximizes rumor mitigation across various social networks.
    Specifically, we observe a minimum improvement of 19.1\% over suboptimal solutions.
    Notably, as the size of the social network increases, the performance of our approach in  rumor mitigation further improves, achieving an over 37\% reduction in rumor spread by blocking only 10\% of the edges, and the improvement compared to other baselines rises from 19\% to 25\%. 
\end{itemize}

\begin{table}[t]
\small
\caption{Rumor mitigation performance comparison.}
\vspace{-10px}
\label{tab::overall}
\begin{tabular}{c|ccc|c}
\toprule
\multirow{2}{*}{\textbf{Method}} & \multicolumn{3}{c|}{\textbf{Twitter}} & {\textbf{Facebook}} \\
& \textbf{S} & \textbf{M} & \textbf{L}& \textbf{L}\\
\midrule
HSD & 16.12 & 13.48 & 12.93 & 13.34 \\
HSC & 18.96 & 20.95 & 18.65 & 25.00  \\
\hline
GA & 17.32 & 14.37 & 12.77 & 11.33 \\
SA & 19.30 & 13.93 & 13.34 & 10.98 \\
\hline
PR & 16.56 & 13.92 & 12.50 & 13.62  \\
KED & 19.80 & 22.16 & 17.57 & 18.30 \\
GBP & \uline{28.53} & \uline{26.21} & 24.32 & 28.46 \\
\hline
DRLN & 28.47 & 26.16 & \uline{26.81} & \uline{29.67} \\
DRLE (ours) & \bf{34.81} & \bf{31.22} & \bf{32.18} & \bf{37.26} \\
\hline
impr\% v.s. DRLN & +22.3\% & +19.3\% & +20.1\% & +25.6\% \\
\bottomrule
\end{tabular}
\vspace{-10px}
\end{table}

\begin{figure}[t]
    \centering
    \includegraphics[width=0.95\linewidth]{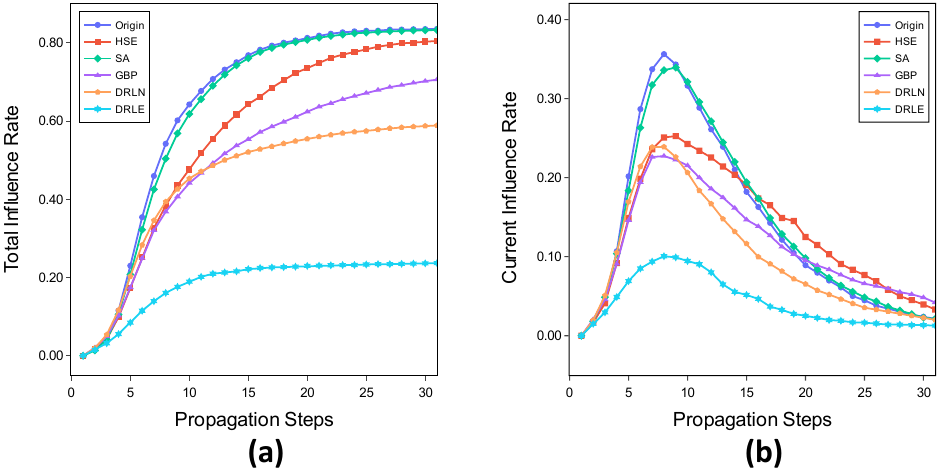}
    \vspace{-10px}
    \caption{
    (a) The cumulative percentage of affected users at each propagation step.
    (b) The proportion of users who are spreading rumors at each propagation step.
    }
    \vspace{-10px}
    \label{fig::dynamic}
\end{figure}

We visualize the rumor propagation at each step in Figure \ref{fig::dynamic}.
In the original network, rumor spreads rapidly upon their emergence, reaching the peak within only 10 steps and affecting more than 60\% of the users in total.
After implementing the proposed rumor mitigation strategy, there is a substantial suppression of rumors.
Specifically, no more than 20\% of users are affected in the first 10 steps of propagation, and the peak impact of rumors also dropped significantly from 35.6\% to 10.1\%.
From a global perspective, our approach reduces the proportion of total users affected by rumors from 83.6\% to 23.7\%, whereas baselines only temporarily block rumors, resulting in at least 58.8\% of users ultimately being affected.

\subsection{Ablation Study}\label{sec::ablation}
We conduct ablation experiments to show the effectiveness of several key designs of the knowledge-informed model.

\begin{figure}[t]
    \centering
    \includegraphics[width=0.95\linewidth]{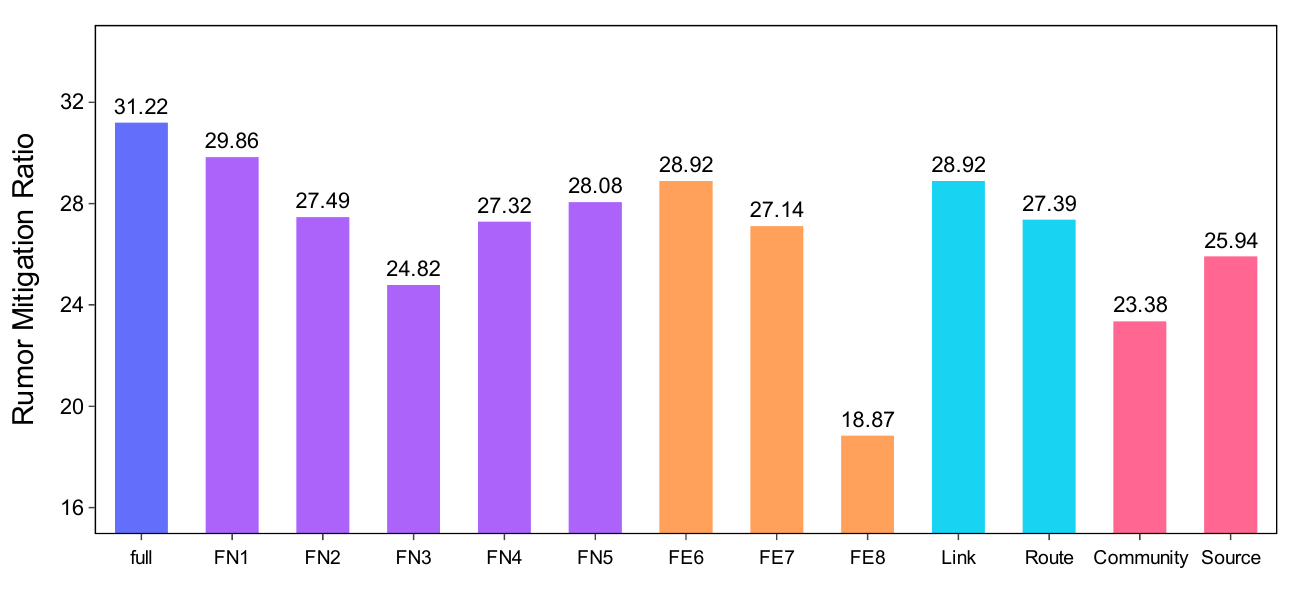}
    \vspace{-15px}
    \caption{
    The result of ablation experiments, which removes features for nodes (FN) and edges (FE), GNN (Link and Route) and the rumor-oriented design in policy network (Community and Source), respectively.
    Best viewed in color.
    }
    \vspace{-15px}
    \label{fig::ablation}
\end{figure}

\noindent{\textbf{Features for Topology and Propagation}}
Figure \ref{fig::ablation} shows the importance of each feature by removing them individually from the well-trained model.
The removal of propagation-related edge centrality (FE8) results in the most significant performance degradation (-39.6\%), and the decline is also significant (-20.5\%) for propagation-related node centrality (FN3), which is reasonable as the propagation-related centralities identify critical edges for rumor propagation.
The carefully designed features for topological structure and propagation can capture the key factor of rumor spreading,  which is critical for rumor mitigation.

\noindent{\textbf{Link-Route-aware Message Passing}}
The proposed GNN model combines the functions of edges as both links and routes in the spread of rumors.
To assess the impact of propagation messages, we conduct experiments by eliminating message passing between nodes and edges, as illustrated in Figure \ref{fig::ablation}.
We can find that the removal of node and edge propagation results in a 7.4\% and 12.3\% reduction in the mitigation ratio, respectively.
This result confirms the advantage of our link-route-aware message passing in aggregating local and global information.

\noindent{\textbf{Rumor-oriented Policy Network}}
As shown in Figure \ref{fig::ablation}, ablation experiments demonstrate the significance of community and rumor source identity in achieving high-quality edge selection.
In specific, when we neglect the correlation with the rumor source, the effectiveness of mitigation decreased by 16.9\%.
Moreover, the decline becomes more pronounced, reaching 25.11\% when we overlook community information.

\vspace{-2px}

\section{Conclusion}

In this paper, we present an RL framework to eliminate the impact of misinformation on social media platforms, with low user awareness and business cost.
We propose a knowledge-informed agent trained with a randomized algorithm, which blocks a minimal set of links to mitigate rumors.
Through extensive experiments conducted on real-world social media platforms, our method demonstrates a substantial effectiveness in curbing the spread of misinformation, showcasing an improvement of over 21.1\% compared to baseline methods. 
In the future, we aim to extend our framework to solve more complicated misinformation challenges, such as combating multiple rumor sources in dynamically changing social networks.

\begin{acks}
This work is supported in part by National Natural Science Foundation of China under grant U23B2030, National Key Research and Development Program of China under grant 2020YFA0711403. This work is also supported in part by Beijing National Research Center for Information Science and Technology (BNRist).
\end{acks}

\bibliographystyle{ACM-Reference-Format}
\balance
\bibliography{references}
\nobalance

\end{document}